\documentclass[12pt]{article} 
\usepackage[dvips]{graphics}
\title{Testing the Alignment Tendency of Some Polarized Radio Sources}  
\author{{\it Richard Shurtleff~}\thanks{affiliation and mailing 
address: Department of Science, 
Wentworth Institute of Technology, 550 Huntington Avenue, 
Boston, MA,  02115, USA, telephone: (617) 989-4338, FAX:  617-989-4010, e-mail: shurtleffr@wit.edu}} 
\begin{document} 
          
\maketitle 

\begin{abstract} 

Measuring the alignment of polarized radio sources requires comparing vectors at different locations on the sky, i.e. on a sphere. A test of alignment is derived herein. While both large scale and coordinate independent, the test avoids the mathematical subtleties involved when comparing vectors at different locations on a curved surface. Applied to 5442 sources drawn from a published catalog, the analysis finds a level of alignment that would be matched by only 7\% to 14\% of data sets with the same sources but with random polarization directions. The locations of the sources involved and the directions that the vectors favor and the regions avoided are described as well.

\vspace{0.5cm}
 
\vspace{0.5cm}

\end{abstract}
\pagebreak
\section{Introduction} \label{into}

When distant objects emit polarized radiation, it is natural to ask if the polarization directions tend to align or, more to the point, whatever  level of alignment is observed, how likely is it that random-angled sources would show the same level of alignment. An investigation of optical sources\cite{Optical} finds evidence for alignment, adding to the motivation to treat radio sources.

The polarized radio sources analyzed here are from a published catalog containing 12746 sources.\cite{J/MNRAS/376/371} Each record contains the location, position angle of the polarization vector, as well as the Stokes parameters and the rms uncertainties of the Stokes parameters. Compared with previous analyses,\cite{J/MNRAS/380/162,Janskys3} different criteria are applied here to sift the data and a different statistical test is used. 

The statistical test, dubbed the `great circle test', is developed herein in detail to measure alignment. It improves a test applied to the optical data.\cite{QSOsShurtleff} Both versions differ from tests that compare directly tangent vectors at different points on a sphere which for some tests involve parallel transport and others treat regions small enough that transport is not important. Instead the great circle test compares one tangent vector at one point with any separate location at a second point on the sky in a geometric way that in essence compares two vectors at one point, the source, and is coordinate independent.

\begin{figure}[ht]  
\centering
\vspace{0cm}
\hspace{0in}
\includegraphics*[0,30][360,330]{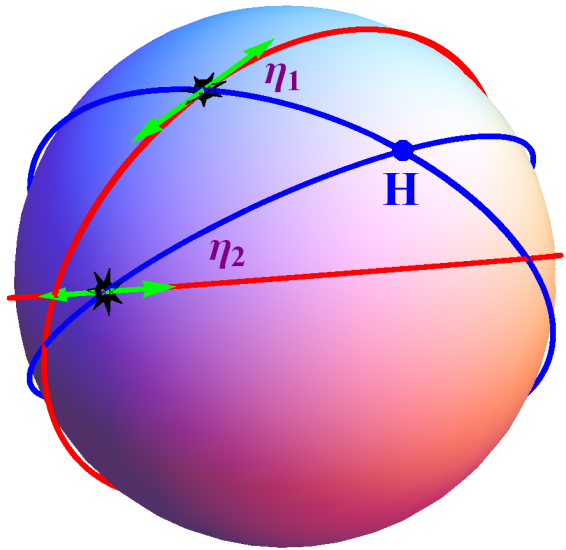}
\caption{(Color online) {\it{The alignment of a point $H$ and a polarization vector, shown here for two sources. }} The angle $\eta$ between the great circle in the direction of a polarization vector and the great circle toward $H$ is a measure of the alignment of $H$ with the vector. The smaller the average angle $\bar{\eta}$ = $(\eta_{1} +\eta_{2})/2$ the more aligned the point $H$ is with the polarization vectors of sources 1 and 2. Some points $H$ are more aligned with the vectors than other points.   }
\label{CelSphere}
\end{figure}

The basic idea can be explained quickly. A single polarization vector pointing out from its source aligns perfectly with every point on the sky directly in front of it, making a great circle that winds completely around the Celestial sphere. Starting at the source, but at some angle $\eta$ with the polarization vector, is another great circle. All the points $H$ on this great circle share the same angle $\eta$ as a measure of their alignment with the polarization vector. See Fig. \ref{CelSphere}.

For a sample of many radio sources, to measure how well a point $H$ on the sky aligns with its many polarization vectors, find the angle $\eta$ for each vector to $H$ and calculate the average angle $\bar{\eta}(H).$  If the polarization vectors tend to align, then there should be some location $H$ that the vectors tend to align with, a place with minimum $\bar{\eta}(H).$  This is the great circle test for alignment. The other extreme, the tendency to point away from a specific location is treated equally, so the test rests on the root-mean-square of the extremes of the average alignment angle $\bar{\eta}_{\mathrm{rms}},$ i.e. the rms of $\bar{\eta}_{\mathrm{min}}$ and $\bar{\eta}_{\mathrm{max}},$ alignment and misalignment.

Larger samples have less chance of alignment than less populated samples. To compare samples with different numbers $N$ of sources, we normalize the rms angle $\bar{\eta}_{\mathrm{rms}}$ to a weighted rms angle $w$ so that the average of $w$ for samples with any number $N$ sources but with randomly chosen polarization angles all give unity for $w,$ no matter what $N$ is. So the great circle test produces a weighted rms angle $w$ for a sample of radio sources and the alignment of a sample is judged on the value of $w.$

Some sources were cut from the catalog before analysis. A few incomplete or suspicious entries were found and cut, but many sources had weak polarization intensity leading to unacceptable uncertainty in the a polarization vector position angle. Sources were cut with an uncertainty greater than $\pm 14^{\circ}.$ Though intensity and uncertainty are related, the effect of the uncertainty cut here most closely  resembles keeping sources with more than 1 mJy of polarization flux. Of the 5442 sources analyzed in this report, 4150 have a flux greater than 1 mJy. We analyze 5442 sources.

Each sample of radio sources analyzed consists of the sources in a $24^{\circ}$ radius circular region centered on a point of a grid with $2^{\circ}$ spacings in longitude (RA) and latitude (dec). Like the entire catalog, the sources are all in the North Equatorial hemisphere, declination $\delta >$ 0. The grid has 5350 points. There are 5350 samples to analyze, with from 139 to 652 sources each. Such numbers require mathematical software for the calculations.\cite{Mathematica} 

Before analyzing the actual data, random polarization position angles were assigned to the sources in 100 sets and these data were analyzed. This produces a baseline of random results which can be used to judge the significance of the weighted rms angle $w.$ Only 1\% of random-angle samples had $w$ greater than 2.25, so any sample of actual observed data with $w$ greater than 2.25 is considered ``very significant.''

Using the polarization angles from the catalog, 158 samples  of the 5350 samples have very significant values of $w.$ Of these 158 very significant samples, a cluster of 90 samples is found near Right Ascension $-119^{\circ}$ and declination $57^{\circ}.$ And there is a 57-sample cluster around (RA,dec) $\approx$ $(+40^{\circ},12^{\circ}).$

The 90-sample cluster has vectors tending to align toward the region around (RA,dec)  $ \approx (29^{\circ},33^{\circ})$ and its diametrically opposite region.  The vectors tend to point away from  (RA,dec)  $ \approx (-50^{\circ},-15^{\circ})$ and its diametrically opposite region. For the cluster of 57 very significant samples, the vectors tend to point toward (RA,dec) $\approx (-133^{\circ},+70^{\circ})$ and its opposite and the vectors point away from $ \approx (-55^{\circ},-4^{\circ})$ and its diametrically opposite location.   

The sky is populated with ten regions of interest: the cluster of sources with its two diametrically opposite regions of alignment and its two regions of avoidance, and double that because there are two clusters. Some coincidences occur, so that the location of these regions may not be arbitrary. 

The four regions of avoidance of the two clusters form two at one pole and two at the opposite pole of a preferred orientation for the Celestial sphere. The remaining six regions loop around the equator of the preferred system with an alignment region of one cluster coincident with the sources of the other cluster.

It is interesting that the analysis leads to a preferred orientation of the sky. But, such coincidences were not expected before-hand and finding something interesting is a likely outcome of any investigation. Thus the coincidences do not count as evidence of polarization vector alignment. 

Including uncertainties in measurements spreads out the results of the analysis. For example, the 158 very significant samples turns out to be at the high end when uncertainties are considered, with from 111 to 158 being the range in values with uncertainties factored in.

For the level of alignment shown by the 5442 radio sources one can take the result that 111 to 158 samples have very significant alignments. Repeating the analysis 100 times, but with random angles in place of the observed polarization position angles, shows that the median number of samples in the random cases is about 32 very significant samples, so the observed data produces many more very significant samples than the median random data. Only about 7\% to 14\% of the random-angled runs have more than 111 to 158 very significant samples.  Since the likelihood is 7\% to 14\% that these results are by chance, the analysis here provides evidence that alignment of the polarization vectors occurs by some physical process or processes.

\section{Configuring the data} \label{data}


The catalog contains  data for the JVAS/CLASS flat-spectrum radio source surveys containing 12746 partially linearly polarized radio sources in the North Equatorial hemisphere. Each record contains the source's Right ascension and Declination (J2000), position angle of its polarization vector, as well as the Stokes parameters I flux, Q flux, and U flux, and their rms uncertainties $\sigma$I, $\sigma$Q, $\sigma$U. A detailed discussion of the data can be found in reference [2].


Many sources were cut for this analysis. Seventy-five incomplete records with missing or null flux uncertainties $\sigma$I, $\sigma$Q, or $\sigma$U and one record with equal I flux and its uncertainty $\sigma$I were removed. Removing 76 records leaves 12670 records. 

The polarization position angle PA, $\theta_{p},$ is listed for each record, but its uncertainty $\sigma_{ \theta}$ is not listed. It is important to know the uncertainty in the position angle when evaluating alignment. Therefore, the uncertainties $\sigma_{ \theta}$ were estimated from the uncertainties  $\sigma_{\mathrm{Q}}$ and  $\sigma_{\mathrm{U}}$ in the Stokes parameters.

The polarization position angle $\theta_{p}$ for a source with observed Q,U-Stokes parameters ($q$,$u$) satisfies  both
\begin{equation} \label{thetaFROMqu}
q  =  p \cos{2\theta_{p}} \quad {\mathrm{and}} \quad u  =  p \sin{2\theta_{p}} \, ,
\end{equation} with $p$  =  $\sqrt{q^2 + u^2} .$  
Plotted in a Q-U rectangular coordinate system, $(q,u)$ are the rectangular coordinates while $(p,2\theta_{p})$ are the polar coordinates for the data point.

\begin{figure}[ht]  
\centering
\vspace{0cm}
\hspace{0in}\includegraphics[0,0][360,360]{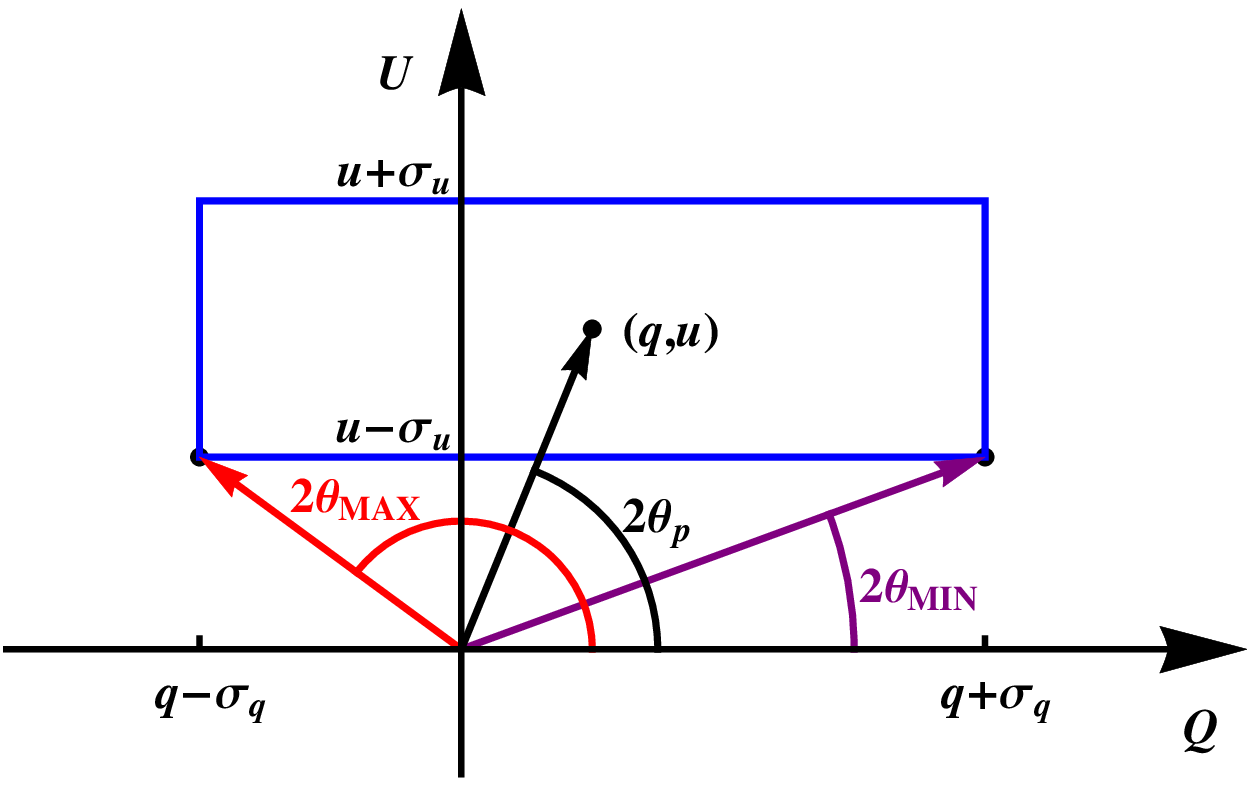}
\caption{(Color online) {\it{The polarization position angle $\theta_{p}$ determined from the Stokes parameters $Q$-flux and $U$-flux.}}  Suppose the measured fluxes are $(q,u)$ with uncertainties $(\sigma_{q},\sigma_{u}),$ here with $\sigma_{q} > $ $q.$ The best value for position angle is the angle $\theta_{p},$ which is half the angle to $(q,u).$  For simplicity, assume that any flux values inside the uncertainty rectangle are consistent with the measurements. The extreme angles that intercept the rectangle determine the  uncertainty $\sigma_{\theta}.$  Here the extreme angles run to the lower two corners of the rectangle giving the max and min values $\theta_{{\mathrm{MAX}}}$ and $ \theta_{{\mathrm{MIN}}}.$ We estimate  the uncertainty to be half the difference, $\sigma_{\theta}$ = $(\theta_{{\mathrm{MAX}}} - \theta_{{\mathrm{MIN}}})/2.$ }
\label{UncertaintyRect}
\end{figure}

A measurement $(q,u)$ together with its uncertainties $\sigma_{q}$ and $\sigma_{u},$ all catalog data, produce an uncertainty rectangle bounded by $q \pm \sigma_{q}$ and $u \pm \sigma_{u}.$ The angle $2\theta_{p}$ is measured to the center of the uncertainty rectangle. There are just three cases: (i) the uncertainty rectangle can be completely in one quadrant, (ii) it can range over two quadrants or (iii) it may have portions in all four quadrants. Figure~(\ref{UncertaintyRect}) shows a two-quadrant case. 

Maximum and minimum angles, $\theta_{{\mathrm{MAX}}}$ and $\theta_{{\mathrm{MIN}}},$ occur in the one or two quadrant cases (i) or (ii). For case (iii) both uncertainties are larger than the magnitude of the value,  $\sigma_{q} > \mid q \mid$ and $\sigma_{u} >\mid u \mid,$ and the uncertainty rectangle includes the origin. For case (iii), the uncertainty range includes all angles, so $\theta_{p}$ is undetermined.  

For sources with maximum and minimum values, $\theta_{{\mathrm{MAX}}}$ and $\theta_{{\mathrm{MIN}}},$ cases (i) and (ii), the range is then halved to give the uncertainty in polarization position angle, $\sigma_{ \theta}.$ We have
\begin{equation} \label{sigTHETA}
\sigma_{ \theta} \equiv \frac{1}{2} \left( \theta_{{\mathrm{MAX}}} - \theta_{{\mathrm{MIN}}} \right) \, .
\end{equation}
Sources with uncertainties greater than $14^{\circ}$, $\sigma_{ \theta} \geq 14^{\circ},$ are cut. The choice of $14^{\circ}$ is consistent with the cut for polarized optical sources treated in earlier work.\cite{Optical} After the cuts, 5442 sources remain to analyze.


To construct samples of radio sources to analyze, a grid of points in the Northern hemisphere was set up separated by two degrees in RA and two degrees in dec. 
The grid consists of the 5350 points (RA,dec) = $(\alpha_{i},\delta_{j})$ determined by
\begin{equation} \label{grid}
 \alpha_{i} = \frac{2^{\circ} \, i}{(\cos{(\delta_{j})}+0.01)} \quad ; \quad \delta_{j} = 2^{\circ} \, j \, ,
\end{equation} 
where the RA index $i \in$ $\{0, ...,$  Ceiling$\left[180 (\cos{(\delta_{j})} + 0.01)\right]\}$ and the dec index $j \in$ $\{ 0,1,2,$ $ \ldots , 45\}.$ The cosine divisor compensates for decreasing radius of circles parallel to the equator as declination increases.  The term 0.01 added to $\cos{\delta}$  prevents infinity making an appearance as $\delta$ approaches zero near the pole.

The samples analyzed consist of all radio sources within $24^{\circ}$-radii circular regions centered on the grid points.  At the celestial equator the regions are effectively semicircular, since there are no southern hemisphere sources. However lopsided the distribution of sources in a circular region, all results for a sample are anchored to the grid point at its center.

The radius chosen for this paper, $24^{\circ},$ was selected because $30^{\circ}$ samples produce a weak, washed out effect. Smaller values, $6^{\circ}$, $12^{\circ}$, $18^{\circ}$ were tried, and produce more detail that may be analyzed elsewhere. 

The 5350 samples analyzed consist of the sources found in the $24^{\circ}$-radius circular regions centered on a $2^{\circ}$-spaced grid in the Northern Equatorial Hemisphere. Each source has a data record with the source's RA and dec, Stokes parameters I,Q,U and their uncertainties, and the polarization position angle and its uncertainty.

\pagebreak

    \section{Great Circle Test for Alignment} \label{GCT}

The test relies on the notion of alignment between a polarization vector at a source located on the Celestial sphere and some other point on the Celestial sphere, i.e. the sky. See Fig. \ref{CelSphere}. If the polarization vectors of many sources in a sample align with the same point, then the polarization vectors themselves are aligned.

A vector $\overrightarrow{V}$ tangent to a sphere at a point $O$ on the sphere determines a great circle. A second point $H$ on the sphere together with point $O$ determines another great circle. The two great circles intersect at $O$ and $-O$ and their tangents at $O$ determine an acute angle $\eta.$ The smaller the angle $\eta,$ the more aligned the point $H$ is with the vector $\overrightarrow{V}$ at $O.$ The angle $\eta$ gauges the alignment between a polarization vector $\overrightarrow{V}$ at a source located on the sky and some other point $H$ also located on the sky.

Given the point $H$ and a set of $N$ vectors $\overrightarrow{V}_{i}$ with $i \in$ $\{1,2,...,N\},$ at points $O_{i} \neq$ $H$ on the sphere, the smaller the average angle 
\begin{equation}  \label{aveETAeq}
\bar{\eta}(H) = \frac{1}{N}\sum_{i=1}^{N}{\eta_{i}(H)} \, ,
\end{equation}
 the more the set of vectors tends to align with $H.$ A multipole expansion in spherical harmonics is a useful way to represent the function  $\bar{\eta}(H).$ 

Given the set of $N$ vectors, the average angle $\bar{\eta}(H)$ is a function defined at all the points $H$ in the sky, except at the $N$ points $O_{i},$ since determining the great circle $O_{i}H$ requires two distinct points. Since all great circles through $H$ also contain the diametrically opposite point $-H$, one sees that  $\bar{\eta}(H)$ = $\bar{\eta}(-H).$ Thus $\bar{\eta}(H)$ is an even function with nonzero even order multipoles.


Expanding the function $\bar{\eta}(H)$ in even order multipoles yields
\begin{equation} \label{expand1}
\bar{\eta}(H(\Omega)) = \bar{\eta}(\Omega) =  \bar{\eta}_{0} + \bar{\eta}_{q}(\Omega) + \bar{\eta}_{4}(\Omega) + ... = \sum^{\infty}_{l=0}\sum^{l}_{m=-l}{a^{l}_{m}Y^{l}_{ m}(\Omega)} \; ,
\end{equation}
where $\bar{\eta}_{l}$ are the even order multipoles indicated by the even integer $l,$  with quadrupole $\bar{\eta}_{q}$ = $\bar{\eta}_{2}$ indicated with the subscript $q.$ The symbol $\Omega$ stands for the spherical coordinates polar angle and longitude,  $\Omega$ = $(\theta,\phi).$ Sometimes it is convenient to use alternates for $\Omega,$ (RA,dec) = $(\alpha,\delta)$ = $(\phi,\pi/2-\theta),$ in radians.  The $Y^{l}_{m}$ are real-valued Tesseral Spherical Harmonics\cite{realForm} which satisfy the orthonormality conditions of the more common complex valued spherical harmonics, but avoid the complexity by rewriting complex phase factors $e^{\pm i m \phi} $ with real valued trig functions $\sin{(m\phi)}$ and $\cos{(m\phi)}.$ 

The coefficients $a^{l}_{m}$ are determined by integration,
\begin{equation} \label{alm}
a^{l}_{m} = \int{\bar{\eta}Y^{l}_{m}} d\Omega \; ,
\end{equation}
where $d\Omega$ = $\sin{\theta} d\theta d\phi$  = $\cos{\delta} d\delta d\alpha \, .$

One must worry about a mathematical detail. The function $\bar{\eta}(H)$ is not defined at the locations of the $N$ sources, a finite number of points on the sphere. This follows because one cannot get a unique great circle through $H$ and the source when $H$ and the source are the same point.

The remedy is from elementary calculus. In a circular neighborhood of radius $\epsilon$ radians centered at each source on the sphere, replace the value of $\bar{\eta}(H)$ throughout with the max possible acute angle, $\bar{\eta}(H)$ = $\pi/2.$ Now $\bar{\eta}(H)$ is defined everywhere and the $N$ circular regions contribute an amount $\delta a^{l}_{m}$ =  $\pi^{2} \epsilon^{2}\sum Y^{l}_{m}/2$ to the integral, with the $Y^{l}_{m}$ evaluated at and summed over the sources. The spherical harmonics $Y^{l}_{m}$ are finite bounded functions. Thus the contribution $\mid \delta a^{l}_{m} \mid$  can be made as small as one wishes by choosing $\epsilon$ small enough. In this way, the integral can be carried out to whatever precision one likes.

For a sample with significant alignment, the monopole $\bar{\eta}_{0}$ represents the constant average value of acute angles $\bar{\eta}$ , $\bar{\eta}_{0}$ = $\pi/4$ radians = $45^{\circ}.$ And the $l$ = 4 and higher multipoles are much smaller than the quadrupole, $l$ = 2.  

Angular coordinates on the sphere can be converted to rectangular coordinates with the definition of the unit vector $\hat{r}$
\begin{equation} \label{rhat}
  \hat{r} = \{\hat{r}_{x},\hat{r}_{y},\hat{r}_{z}\} = (\cos{\alpha} \cos{\delta} ,  \sin{\alpha} \cos{\delta}, \sin{\delta}) \, .
\end{equation}
Clearly $\hat{r}$ is a unit vector; by squaring and adding components, one sees that $\hat{r}\cdot\hat{r}$ = 1.

For illustration and because multipoles of higher order than the quadrupole are insignificant, we list the five real-valued spherical harmonics for the quadrupole $l$ = 2.\cite{realForm} These can be written with either angular coordinates $(\alpha,\delta)$ or rectangular coordinates $\{\hat{r}_{x},\hat{r}_{y},\hat{r}_{z}\},$
\begin{equation} \label{Y2m} Y_{-2}^{2} = - k \sin{\left(2\alpha\right)} \, \cos^{2}{\left(\delta\right)}= - 2 k \hat{r}_{x} \hat{r}_{y}
\end{equation}
$$ Y_{ -1}^{2} = - k \sin{\left(\alpha \right)}\cos{\left(2\delta \right)} = - 2 k \hat{r}_{y} \hat{r}_{z} $$
$$ Y_{0}^{2} = \frac{k}{\sqrt{3}}\left(3\sin^{2}{\left( \delta \right)} - 1 \right) = \frac{k}{\sqrt{3}}\left(3\hat{r}_{z}^{2} - 1 \right)$$
$$ Y_{+1}^{2} = - k \cos{\left(\alpha \right)}\cos{\left(2\delta \right)} =  - 2 k \hat{r}_{x} \hat{r}_{z} $$
$$ Y_{+2}^{2} = +k \cos{\left(2 \alpha \right)} \cos^{2}{\left(\delta \right)} = k\left(\hat{r}_{x}^{2} -\hat{r}_{y}^{2}\right) \; ,$$
where $ k = \sqrt{15/\left(16 \pi \right)} .$

The functions $Y_{m}^{2}$ are orthogonal to all non-quadrupoles, $l \neq 2,$ and are orthonormal among themselves. This gives (\ref{alm}) from (\ref{expand1}). With $m^{\prime},m \in$ $\{-2,-1,0,1,2\},$ one can show that
\begin{equation} \label{YYortho}
 \int_{\alpha = -\pi}^{\pi}\int_{\delta = -\pi/2}^{\pi/2}{Y_{ m^{\prime}}^{2}Y_{m}^{2} \, d \Omega} = \delta^{m^{\prime} m} \; ,  
\end{equation}
where the Kronecker delta $ \delta^{m^{\prime} m}$ is one when the $m^{\prime}$ and $m$ are equal and zero otherwise.

By (\ref{expand1}) and (\ref{Y2m}),  it follows that the quadrupole $\bar{\eta}_{q}$ is quadratic in the components of $\hat{r}.$ We have
\begin{equation} \label{etaq2}
\bar{\eta}_{q} = \sum_{m = -2}^{+2}{a_{m} Y_{ m}^{2} } = \hat{r}\cdot T \cdot \hat{r} \; ,
\end{equation} 
where, by using the expressions (\ref{Y2m}) for $Y_{m}^{2}$ in terms of $\hat{r},$ we see that
\begin{equation} \label{Tensor}
  T = \frac{1}{4}\sqrt{\frac{15}{\pi}}\pmatrix{a_{+2}-\frac{1}{\sqrt{3}} a_{0} &  a_{-2} &  a_{+1} \cr  a_{-2} & -a_{+2}-\frac{1}{\sqrt{3}} a_{0} &  a_{-1} \cr  a_{+1} &  a_{-1} & \frac{2}{\sqrt{3}}\,a_{0} } \, .
\end{equation}
Thus it is straightforward to go back and forth between the components of the matrix $T$ and the quadrupole coefficients $a_{m}.$ Note that $T$ is both symmetric and traceless.\cite{STTensor}

Because $\bar{\eta}_{q}$ is a scalar under rotations, $\bar{\eta}_{q}$ = $\bar{\eta}_{q}^{\, \prime},$ it follows from (\ref{etaq2}) that $T$ transforms as a 2nd rank tensor under rotations. This makes $T$ convenient when dealing with rotations.

One can also use $T$ to find the maxima and minima of the quadrupole $\bar{\eta}_{q}.$ The extreme values of $\bar{\eta}_{q}(H)$ = $\hat{r}\cdot T \cdot \hat{r}$ occur where its gradient vanishes. While one can consider what three dimensional harmonic function has the values $\bar{\eta}_{q}(H)$ on the unit sphere, it is perhaps easier to stay on-sphere and constrain the coordinates $(\hat{r}_{x},\hat{r}_{y},\hat{r}_{z})$ to remain a unit vector. The sum of the components squared must remain unity to keep $\hat{r}$ on the sphere. 

With a Lagrange multiplier to handle the constraint, we seek the extrema of the function
\begin{equation} \label{f}
f(\overrightarrow{r}) = \bar{\eta}_{q}(H) - \lambda \overrightarrow{r} \cdot \overrightarrow{r}  = \overrightarrow{r} \cdot T \cdot \overrightarrow{r}  - \lambda \overrightarrow{r} \cdot \overrightarrow{r}  \, ,
\end{equation}
where $\lambda$ is the Lagrange multiplier, whose allowed values are to be determined later. The vector $\overrightarrow{r}$ is not necessarily a unit vector, because the components need to vary independently one from the others. So, the hat is dropped.

The extrema occur when the gradient of $f$ vanishes,
\begin{equation} \label{delf}
\overrightarrow{\bigtriangledown} f = \overrightarrow{\bigtriangledown}{\left(\overrightarrow{r}\cdot T \cdot \overrightarrow{r}  - \lambda \overrightarrow{r} \cdot \overrightarrow{r}\right)} = 2\left( T \cdot \overrightarrow{r}  - \lambda \overrightarrow{r} \right) = 0 \, ,
\end{equation}
where the gradient is $\overrightarrow{\bigtriangledown}$ = $\{\partial/\partial x,\partial/\partial y, \partial/\partial z \},$ with $\{x,y,z\}$ the three components of $\overrightarrow{r},$  and the 0 is actually $(0,0,0).$ This reroutes the search for extrema to the eigenvalue problem, 
\begin{equation} \label{eigenPROBLEM}
T \cdot \hat{r} = \lambda \hat{r} \, ,
\end{equation}
for the real symmetric tensor $T,$  a topic covered in basic quantum mechanics. The eigenvectors have arbitrary nonnegative magnitudes, so the vector $\overrightarrow{r}$ can be required to be a unit vector $\hat{r}.$  The Lagrange multipliers have become also eigenvalues of $T.$

Real symmetric matrices like $T$ have real eigenvalues\cite{eigenvalues} and the solutions can be sorted by eigenvalue;
\begin{equation} \label{eigen}
(\lambda_{\mathrm{min}},\pm \hat{r}_{\mathrm{min}})\quad ; \quad (\lambda_{\mathrm{mid}},\pm \hat{r}_{\mathrm{mid}}) \quad ; \quad (\lambda_{\mathrm{max}},\pm \hat{r}_{\mathrm{max}}) \, ,
\end{equation}
where $\lambda_{\mathrm{min}} \leq \lambda_{\mathrm{mid}} \leq \lambda_{\mathrm{max}}.$ For the applications here, the eigenvalues are distinct.

Now, by the definition of $f$ in (\ref{f}), the function $f$ vanishes for any solution of the eigenvalue problem, $f(\hat{r}_{j})$ = $\hat{r}_{j} \cdot T \cdot \hat{r}_{j}  - \lambda \hat{r}_{j} \cdot \hat{r}_{j}$ = $\hat{r}_{j} \cdot (T \cdot \hat{r}_{j}  - \lambda \hat{r}_{j})$ = 0, whenever $T \cdot \hat{r}$ = $\lambda \hat{r},$ i.e. for $j \in$ $\{\mathrm{min},\mathrm{mid},\mathrm{max} \}.$ Thus the max and min eigenvalues, $\lambda_{\mathrm{max}}$ and $\lambda_{\mathrm{min}},$ are max and min values of $\bar{\eta}_{q}(H),$
\begin{equation} \label{maxMINlambdaETA}
\bar{\eta}_{q \,\mathrm{max}} = \left(\hat{r} \cdot T \cdot \hat{r}\right)_{\mathrm{max}}  = \lambda_{\mathrm{max}} \hat{r} \cdot \hat{r} =  \lambda_{\mathrm{max}} \, ,
\end{equation}
because $\hat{r}$ is a unit vector, $ \hat{r} \cdot \hat{r}$ = 1. Similar reasoning goes through for the minimum value of $\bar{\eta}_{q}(H).$ We identify the max-min Lagrange multipliers as max-min values of the quadrupole,
\begin{equation} \label{maxMINlambdaETA1}
\bar{\eta}_{q \,\mathrm{min}} =  \lambda_{\mathrm{min}} \quad ; \quad \bar{\eta}_{q \,\mathrm{max}} =  \lambda_{\mathrm{max}}  \, .
\end{equation}
These are the extreme values for the quadrupole of $\bar{\eta}(H)$ that gauge the strength of the tendency for the sources of a sample to align one with another.

While alignment may be an indicator of the existence of some mutual effect, the underlying process may be more the opposite. Perhaps radio waves are absorbed and scattered on their journey from source to detector in a polarization-dependent process. In that case polarization vectors would seem to avoid the direction scattered the most and by attrition seem to align with the perpendicular direction. In fairness, both the tendency to align with some point on the sky and the tendency to avoid some point on the sky are given equal standing in this paper as indicators of coordinated behavior.

Therefore, treating equally alignment, i.e. $\bar{\eta}_{q \,\mathrm{min}},$ and avoidance, i.e. $\bar{\eta}_{q \,\mathrm{max}},$ suggests the root-mean-square of these extremes, the rms angle
\begin{equation} \label{etaRMS}
\bar{\eta}_{q \,\mathrm{rms}} =  \sqrt{ \frac{\bar{\eta}_{q \,\mathrm{min}}^{2}+\bar{\eta}_{q \,\mathrm{max}}^{2}}{2}}  \, .
\end{equation}
Looking at the calculated values with actual data suggests that the magnitudes $\mid \bar{\eta}_{q \,\mathrm{min}} \mid $ and $\bar{\eta}_{q \,\mathrm{max}}$ are both large if one is, so the distinction is, for most samples, irrelevant. 

The quantity $\bar{\eta}_{q \,\mathrm{rms}}$ could be used to judge the alignment of a sample of radio sources. An adjustment is needed, however, because samples with more sources tend to have smaller values of $\bar{\eta}_{q \,\mathrm{rms}},$ as shown by considering random-angle sources. For a sample of $N$ sources with an rms angle $\bar{\eta}_{q \,\mathrm{rms}},$ we divide by the average $\bar{\eta}^{\mathrm{AVE}}_{q \,\mathrm{rms}}(N)$ over many random-angled samples with the same number of radio sources,
\begin{equation} \label{w}
w = \frac{\bar{\eta}_{q \,\mathrm{rms}}}{\bar{\eta}^{\mathrm{AVE}}_{q \,\mathrm{rms}}(N)}   \, ,
\end{equation}
where $N$ is the number of sources in the sample.  The function $w$ is `weighted' to account for the number of sources. The function $\bar{\eta}^{\mathrm{AVE}}_{q \,\mathrm{rms}}(N)$ is determined in Sec. 4. By definition, $w$ is normalized so that $w$ = 1 for the average random-angle sample with any number of sources.

It is interesting to consider the points on the sky that the polarization vectors favor the most and those points favored least. These are in the directions of the eigenvectors of the  max and min eigenvalues of the matrix $T.$ Since the negative of an eigenvector is an eigenvector with the same eigenvalue, an eigenvector is determined only up to sign, and these points come in pairs of diametrically opposite places on the sky.   

The maximum value of $\bar{\eta}_{q}(H)$ is $\bar{\eta}_{q\,{\mathrm{max}} }$ = $\lambda_{\max}$ which occurs in the direction of the eigenvector $\pm \hat{r}_{\mathrm{max}}.$ And the minimum $\bar{\eta}_{q\,{\mathrm{min}} },$ i.e. $\lambda_{\min},$ occurs for the eigenvector $\pm \hat{r}_{\mathrm{min}}.$ For a single source at $O$ or a sample of sources at various points $O_{i}$ but confined to a small region, we find that the eigenvector $ \hat{r}_{\mathrm{mid}}$ of the mid-range eigenvalue $\lambda_{\min},$ points in the general direction of the source(s). 

One can show that, for distinct eigenvalues, the eigenvectors are orthogonal,\cite{eigenvalues}
\begin{equation} \label{perp}
 \hat{r}_{\mathrm{min}} \cdot \hat{r}_{\mathrm{mid}} = \hat{r}_{\mathrm{min}} \cdot \hat{r}_{\mathrm{max}} = \hat{r}_{\mathrm{mid}} \cdot \hat{r}_{\mathrm{max}} = 0 \, ,
\end{equation}
so the three vectors are the basis vectors of a rectangular coordinate system. This means that the direction of the sources which is roughly $\hat{r}_{\mathrm{mid}}$, the direction of alignment $\hat{r}_{\mathrm{min}}$ and the direction $\hat{r}_{\mathrm{max}}$ avoided by the vectors  $\overrightarrow{V}_{i}$  form a rectangular coordinate system. 

\begin{figure}[ht]  
\centering
\vspace{0cm}
\hspace{0in}\includegraphics[0,0][360,360]{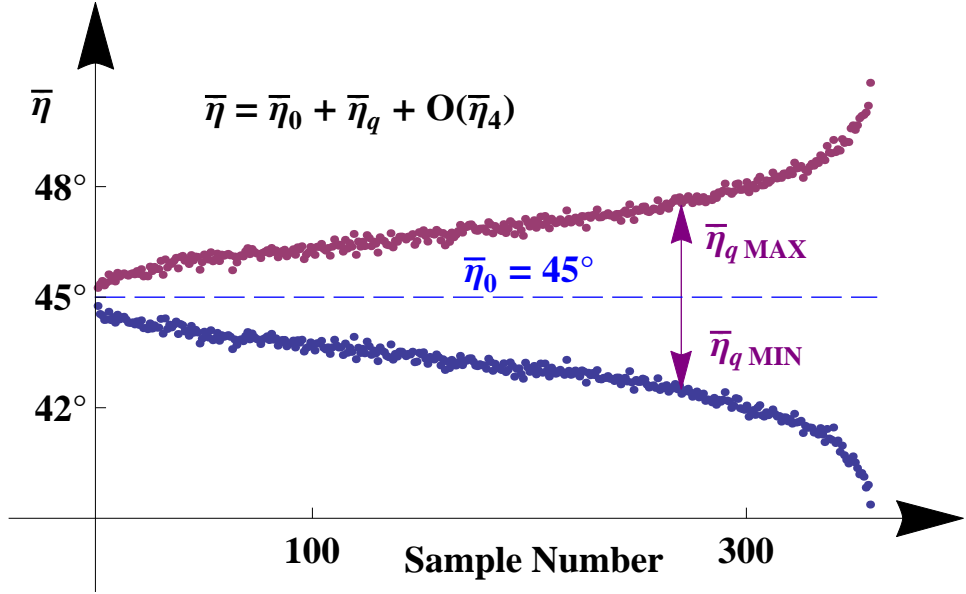}
\caption{(Color online) {\it{Multipole expansion of the polarization angle function $\bar{\eta}(H)$ for some samples with random polarization angles.}} As the average of $\eta,$ Fig. \ref{CelSphere}, over all sources in a sample,  the angle $\bar{\eta}(H)$ varies between the upper and lower arms of the plot as $H$ varies across the sky. Expanded in even order spherical harmonics, $\bar{\eta}(H)$ is sufficiently well approximated by the constant monopole, $\bar{\eta}_{0}(H)$ = $45^{\circ},$  and the quadrupole $\bar{\eta}_{q}(H).$ The max-min values of $\bar{\eta}(H)$ occur at points $H$ on the sky where the max-min values of the quadrupole, $\bar{\eta}_{q \, {\mathrm{max}}}$ and $\bar{\eta}_{q\, {\mathrm{min}}},$ are located. Note the symmetry, $\bar{\eta}_{q \, {\mathrm{max}}} \approx$  $\mid \bar{\eta}_{q\, {\mathrm{min}}}\mid .$   }
\label{EtaMeanings}
\end{figure}

In this special coordinate system, the tensor $T$ is diagonal. Since $T$ is traceless in any coordinate system, it follows that the sum of the eigenvalues vanishes, 
\begin{equation} \label{traceless}
\lambda_{\min} + \lambda_{\mathrm{mid}} + \lambda_{\max} = \bar{\eta}_{q \,\mathrm{min}} + \bar{\eta}_{q \,\mathrm{mid}} + \bar{\eta}_{q \,\mathrm{max}} = 0 \, .
\end{equation}
The values of $\bar{\eta}_{q \,\mathrm{min}}$ and $\bar{\eta}_{q \,\mathrm{max}}$ plotted in Fig. \ref{EtaMeanings} are almost equally above and below the dashed line, so the middle value  $\bar{\eta}_{q \,\mathrm{mid}}$ is small compared to either $\mid \bar{\eta}_{q \,\mathrm{min}}\mid$ or $\bar{\eta}_{q \,\mathrm{max}}.$

In this paper the point $O$ is the location of a radio source with vector $\overrightarrow{V}$ as its polarization vector. Next some formulas are derived to assist in the analysis of the radio source data.

All points on any great circle through the North pole line up neatly in spherical coordinates, ready for calculation. Put a single source at the pole.

Let the direction of the polarization vector $\overrightarrow{V}$ have an RA, $\alpha_{p}.$ The great circle from the pole to any point $H$ in the sky, but not at the pole itself, is a second line with a second possibly different RA, $\alpha_{H}.$ The angle $\eta$ between the two great circles is just the difference in RAs, possibly adjusted by some multiple of $\pi$ to make an acute angle,  $0 \leq \eta \leq \pi/2,$
\begin{equation} \label{NPeta}
\eta =\min{\{|\alpha_{H}- \alpha_{p}|,|\pi-(\alpha_{H} - \alpha_{p}  )|,|2\pi-(\alpha_{H} - \alpha_{p}  )|,...},\} \, , 
\end{equation}
in radians.

With just one source, the average $\bar{\eta},$ (\ref{aveETAeq}), is just $\eta$ and, by the definition of the $Y^{2}_{m}$ in (\ref{Y2m}) and by the values of $\eta$ in (\ref{NPeta}), by the integral in (\ref{alm}), $a^{l}_{m}$ = $\int{\bar{\eta}Y^{l}_{m}} d\Omega,$ with $l$ = 2, we get the quadrupole coefficients $a_{m},$ 
\begin{equation} \label{a2mONE}
a_{-2} = -2 \sqrt{\frac{5}{3 \pi}}\sin{(2\alpha_{p})} \quad ; \quad
a_{-1} = a_{0} = a_{+1} = 0 \quad ; \quad a_{+2} = -2 \sqrt{\frac{5}{3 \pi}}\cos{(2\alpha_{p})} \; ,
\end{equation}
where we note again that the $l$ = 2 is understood and dropped: $a_{m}$ = $a^{2}_{m}.$ One sees that putting the source at the North pole has successfully found formulas.

Away from the North pole, for a single radio source in some sample, it is convenient to have formulas for the quadrupole coefficients $a_{m }.$ The coefficients should depend on the location (RA,dec) = $(\alpha,\delta)$ of the source and its polarization position angle $\theta_{p} \,.$ To get the coefficients $a_{m}$, rotate the known coefficients $a_{m}$ at the pole in (\ref{a2mONE}) to the location (RA,dec) = $(\alpha,\delta).$ 

Begin by specifying the coordinate system. The sky is a sphere of unit radius, unit unspecified, centered on the Earth or Sun. The sources are so far away that either will do. The $z$-axis, the North pole, is in the direction of the Earth's rotation axis and has rectangular coordinates $(0,0,1). $ The $x$-axis is in the direction of the Vernal equinox with coordinates $(1,0,0). $ The $xyz$-coordinate system is right-handed, which determines $y.$

First get the declination correct by rotating about the $y$-axis through an angle $\pi/2 - \delta$ and then rotate by angle $\alpha$ about the $z$-axis to get the correct RA, i.e. longitude. One finds the rotation to do this is
\begin{equation} \label{ROTmatrix}
 R(\alpha,\delta) = \pmatrix{\cos{\alpha} \sin{\delta} & -\sin{\alpha} & \cos{\alpha} \cos{\delta} \cr \sin{\alpha} \sin{\delta} & \cos{\alpha} & \cos{\delta} \sin{\alpha}  \cr -\cos{\delta} & 0 & \sin{\delta} }
 \end{equation}
 One can check that rotating the unit vector $\hat{z}$ = $(0,0,1)$ for the direction of the North pole from the origin gives $R\cdot\hat{z}$ = $(\cos{\alpha} \cos{\delta} ,\cos{\delta} \sin{\alpha}, \sin{\delta}),$ which is correct for the point in the sky with (RA,dec) = $(\alpha,\delta).$  The rotation $R$ takes the $\hat{x}$ unit vector, $(1,0,0),$ to local South, 
\begin{equation} \label{South}
 R\cdot\hat{x} = (\cos{\alpha} \sin{\delta} ,\sin{\alpha} \sin{\delta}, -\cos{\delta}) \, .
\end{equation}
Rotating $\hat{y}$ gives local East at the point $(\alpha,\delta),$
\begin{equation} \label{East}
 R\cdot\hat{y} = ( -\sin{\alpha} ,\cos{\alpha},0) \, .
\end{equation}
With these new directions, we must see what happens to the polarization vector when it moves from the pole.

The polarization angle $\alpha_{p}$ for the source at the North pole is measured, just as RA is measured, counterclockwise from the $x$-axis, looking down on the pole from above the sphere. Thus ``counterclockwise from the $x$-axis looking down on the pole'' rotates into ``counterclockwise from local South with local East to the right.

Since the catalogue lists the polarization position angle $\theta_{p}$ measured clockwise from local North with East to the right, and $\alpha_{p}$ is counterclockwise from due South with East to the right, it follows that  
\begin{equation} \label{Supplementary}
 \alpha_{p} + \theta_{p} = \pi \, , 
\end{equation}
so the two angles are supplementary, together covering the angle from North to South via East.

As mentioned previously, the quadrupole coefficients $a_{m}$ form a tensor $T,$ (\ref{Tensor}). To find the quadrupole coefficients after a rotation, rotate the tensor and deduce the rotated coefficients $a_{m}$ from the rotated tensor's components $T^{ij}.$ 

By the dependence of $T$ on the $a_{m}$ in (\ref{Tensor}), with the single source at the North pole having the coefficients $a_{m}$ in (\ref{a2mONE}), one finds that the tensor $T_{0}$ for the single source at the pole is given by
\begin{equation} \label{T0}
T_{0} =   \frac{5}{2\pi}\frac{1}{\sqrt{3}}\pmatrix{\cos{2\alpha_{p}} &  \sin{2\alpha_{p}} & 0 \cr  \sin{2\alpha_{p}} &  -\cos{2\alpha_{p}} &  0 \cr  0 &  0 & 0 } \, .
\end{equation}
Rotating with $R$ transforms $T_{0}$ to the tensor $T,$
\begin{equation} \label{Tensor2}
 T = R\cdot T_{0}\cdot R^{-1} \, .  
\end{equation}
The matrix is unwieldy, but can be pieced together from the equations for $a_{m}$ below. 

Having rotated the tensor, we can identify the rotated quadrupole coefficients by Eq.~(\ref{Tensor}) keeping in mind by (\ref{Supplementary}) that the polarization angle in the catalog is $\theta_{p}$ = $\pi-\alpha_{p}.$ One finds formulas for the $a_{m}$ of a single source,
\begin{equation} \label{am}
 a_{-2} = +\frac{1}{2} \, \sqrt{\frac{5}{3 \pi}} \, \left[\left(\cos{2\delta}-3 \right)\cos{2\theta_{p}}\sin{2\alpha}+4\cos{2\alpha}\sin{\delta}\sin{2\theta_{p}}  \right]
\end{equation}
$$a_{-1} = +2 \, \sqrt{\frac{5}{3 \pi}} \, \cos{\delta}\left( \cos{2\theta_{p}} \sin{\alpha} \sin{\delta} - \cos{\alpha}\sin{2\theta_{p}}  \right)
$$
$$a_{0} = - \sqrt{\frac{5}{ \pi}}  \, \cos^{2}{\delta} \cos{2\theta_{p}} 
$$
$$a_{+1} = +2 \, \sqrt{\frac{5}{3 \pi}} \, \cos{\delta}\left( \cos{2\theta_{p}} \cos{\alpha} \sin{\delta} + \sin{\alpha}\sin{2\theta_{p}}  \right)
$$
$$ a_{+2} = +\frac{1}{2} \, \sqrt{\frac{5}{3 \pi}} \, \left[ \left( \cos{2\delta}-3 \right)\cos{2\theta_{p}}\cos{2\alpha} - 4\sin{2\alpha}\sin{\delta}\sin{2\theta_{p}}  \right]
$$
These formulas give the quadrupole coefficients $a_{m}$ of the function $\eta(H)$ for a single source located at (RA,dec) = $(\alpha,\delta)$ that has a polarization vector with position angle $\theta_{p}.$

Now consider a sample with $N$ sources scattered about in some way. Since both the arithmetic mean, i.e. the average $\bar{\eta}$ = $\sum{\eta_{i}}/N,$ and the integral for the coefficients, $a^{l}_{m}$ = $\int{\bar{\eta}Y^{l}_{m}} d\Omega,$ are linear in $\eta_{i},$ the coefficients $\bar{a}_{m}$ for the sample with $N$ sources are the averages of the coefficients of the individual sources,
 \begin{equation} \label{a2mMANY}
\bar{a}_{m} = \frac{1}{N}\sum_{i=1}^{N} a_{m\,i} \, ,
\end{equation}
with each $ a_{m\,i}$ determined by substituting the data for the $i$th source into the above formulas for $a_{m}$ in (\ref{am}).
Thus the quadrupole coefficients of a sample can be determined from catalog data, the location and polarization position angle of its sources.

To look for alignment, (i) apply (\ref{am}) to find the quadrupole coefficients $a_{m}$ for each of the 5442 sources remaining after the cuts in Sec. 2. (ii) By (\ref{a2mMANY}) average the coefficients $a_{m}$ for each of the 5350 samples described in Sec. 2.  (iii) Build 5350 tensors $T$ with the average coefficients $\bar{a}_{m}$ and solve the eigensystems of the $T$s for eigenvalues and eigenvectors. The root-mean-square of the pair of max and min eigenvalues is the rms angle $\bar{\eta}_{q \,\mathrm{rms}}$ for the sample. (iv) Next weight each of the 5350 rms angles to account for the different numbers $N$ of sources,yielding 5350 weights $w,$ $w$ = $\bar{\eta}_{q \,\mathrm{rms}}/\bar{\eta}^{\mathrm{AVE}}_{q \,\mathrm{rms}}(N).$

When done, the steps give 5350 samples each with a number $w$ that gauges the alignment of its polarization vectors. The samples with larger weighted rms angles $w$ are better aligned than samples with lower values. 

Beyond finding a measure of the alignments of the various samples, one can use the eigensystem of $T$ for a sample to tell us where the polarization vectors tend to point and where they tend to avoid.

Before applying the great circle test to the data, it is important to estimate significance. How likely is it that similar data, but with random-angles in place of the observed polarization angles would produce samples with as good or better alignments than the observed results? We see next.

\section{Random-angles and Significance} \label{RanSig}

It is important to run the great circle test on random-angled samples in order to judge the significance of results with real data. As discussed in Sec. \ref{data}, the original catalog was cut to 5442 sources. The radio sources were given random-angles between $0^{\circ}$ and $180^{\circ}$ using a pseudo-random generator,\cite{Mathematica} in place of their observed polarization position angles. The location of the sources in the sky was preserved. In this way, one hundred sets of random data were created to analyze. 

As detailed in Sec. \ref{data}, the sources were then organized into 5350 samples each with the sources contained in $24^{\circ}$-radius circular regions centered on a $2^{\circ}$-spaced grid in the Northern Equatorial Hemisphere. In all 100 versions of random data, there were a total of 535000 samples with random-angle data. 

Using the random-angle data, the procedure in the previous section gives the rms angle $\bar{\eta}_{q \,\mathrm{rms}}$ for each of the 535000 samples. The number of sources $N$ in the samples is found to vary from 139 to 652, skipping five values along the way. It follows that there are about a thousand samples for each of the 508 sample sizes $N.$ Averaging the thousand or so values of $\bar{\eta}_{q \,\mathrm{rms}}$ for each $N_{i},$ gives the average rms angle $\bar{\eta}^{\mathrm{AVE}}_{q \,\mathrm{rms}}(N_{i})$ for the 508 values of $N_{i}.$



Since the lowest number of sources is $N$ = 139, which is a long way from zero, the rms angle $\bar{\eta}_{q \,\mathrm{rms}}$ was derived for a single source. By (\ref{maxMINlambdaETA}) and (\ref{maxMINlambdaETA1}), the eigensystem of tensor $T_{0}$ in (\ref{T0}) gives $\bar{\eta}_{q \,\mathrm{rms}}$ = $45.6^{\circ}$ for $N$ = 1. Adding the single source to the others gives 509 rms angles $\bar{\eta}^{\mathrm{AVE}}_{q \,\mathrm{rms}}(N_{i}).$ 

\begin{figure}[ht]  
\centering
\vspace{0cm}
\hspace{0in}\includegraphics[0,0][360,360]{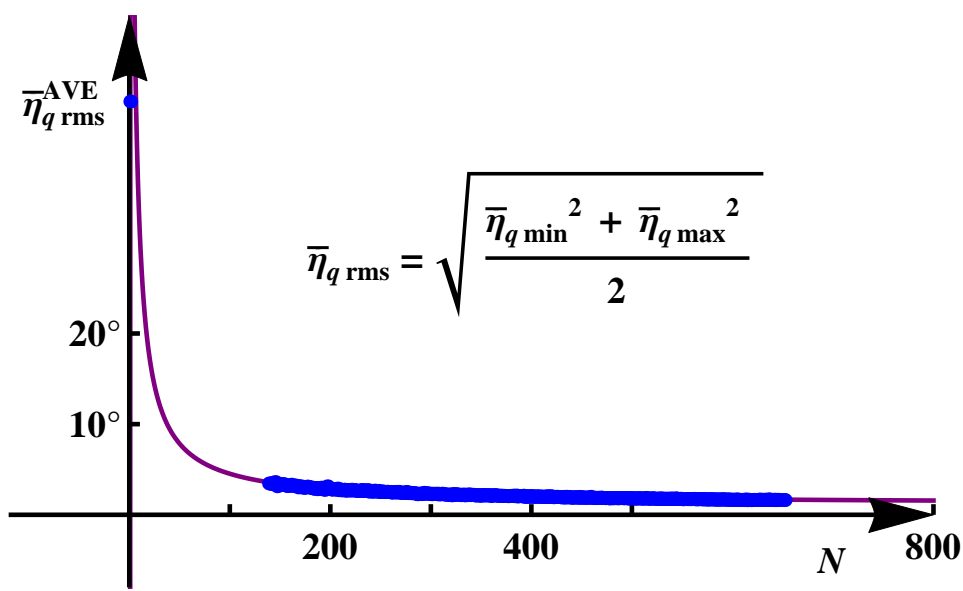}
\caption{(Color online) {\it{Dependence on the number of sources.}} The solid line is a fit to the root-mean-square of the extreme quadrupole values, $\bar{\eta}_{q \, {\mathrm{max}}}$ and $\bar{\eta}_{q\, {\mathrm{min}}}$ in Fig. \ref{EtaMeanings}, for random-polarization-angle samples. For each value of the number $N$ of sources per sample, the average $\bar{\eta}_{q\, {\mathrm{rms}}}$ was found for about a thousand random-angled samples. The samples in this article have $N$ running from $N$ = 139 to 652. To anchor the fit to small $N,$ the value $\bar{\eta}_{q\, {\mathrm{rms}}}$ = $45.6^{\circ}$ at $N$ = 1 for a sample with a single source was included in the fit. The dependence on $N$ is important when comparing samples with different numbers of sources. } 
\label{AveRandomEta(N)}
\end{figure}

A best fit to the now 509 data points was found, yielding 
\begin{equation} \label{etaMINmaxRANfit}
\bar{\eta}^{\mathrm{AVE}}_{q \,\mathrm{rms}}(N) = 1.14704 + \frac{341.119}{N} - \frac{296.671}{N^2} 
\end{equation}
The formula fits the 508 random-angle average rms angles $\bar{\eta}^{\mathrm{AVE}}_{q \,\mathrm{rms}}(N_{i})$ to about 2\%, which is the a root-mean-square fractional difference of the formula (\ref{etaMINmaxRANfit}) and the average random-data value for the 508 sample sizes. See Fig. \ref{AveRandomEta(N)}

Given the average rms angle formula (\ref{etaMINmaxRANfit}) and a sample with $N$ sources and rms angle $\bar{\eta}_{q \,\mathrm{rms}},$ one can now calculate the weighted rms angle 
$w$ = $\bar{\eta}_{q \,\mathrm{rms}}/\bar{\eta}^{\mathrm{AVE}}_{q \,\mathrm{rms}}(N),$ (\ref{w}). Now there are 535000 random-angled samples each with a value of $w.$  

\begin{figure}[ht]  
\centering
\vspace{0cm}
\hspace{0in}\includegraphics[0,0][360,360]{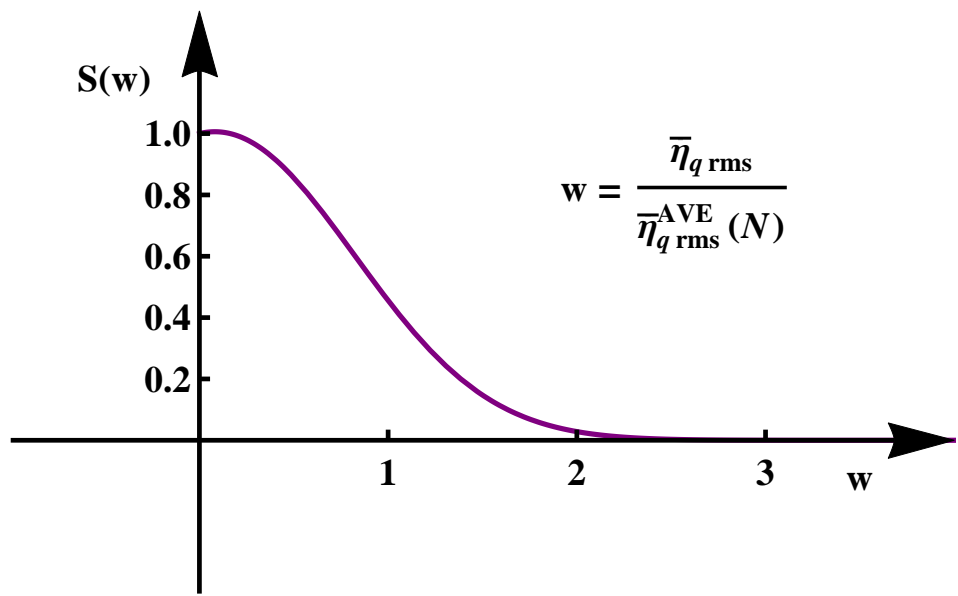}
\caption{(Color online) {\it{Significance of samples of radio sources.}} The significance of a sample is based on the weighted rms angle $w,$ obtained by normalizing the rms angle $\bar{\eta}_{q\, {\mathrm{rms}}}$ to the average of random values, the function $\bar{\eta}^{\mathrm{AVE}}_{q \,\mathrm{rms}}(N)$  in Fig. \ref{AveRandomEta(N)}. Thus for the average random-angle sample, the numerator and denominator in the definition of $w$ are equal, so $w$ is normalized to unity, $w$ = 1, with any $N.$ To find $S(w),$ the weighted rms angle was calculated for 535000 random-angled samples. The fraction of weighted rms angles larger than $w$ is the significance $S(w).$ Thus $S(w_{0})$ = 0.01 means that 1\% of the 535000 random-angled samples had $w$ greater than $w_{0}.$ One finds that $S(w_{0})$ = 0.01 for $w_{0}$ = 2.25, and a sample with an equal or greater $w$ is deemed to have ``very significant'' alignment. }
\label{SigOfw}
\end{figure}

To estimate the likelihood of the alignment of a sample by random-angled data, the 535000 values of $w$ are sorted and the number of $w$s greater than a given $w$ is tallied. Dividing that number by 535000, gives the fraction of samples with weighted rms angle $w$ exceeding the given $w.$ See Fig. \ref{SigOfw}. The function
\begin{equation} \label{S}
S(w) = e^{-0.148469 w-0.902886 w^2-0.0314234 w^3} \, ,
\end{equation}
fits the distribution with a root-mean-square fractional difference of 1.6\%.
The value $S(w_{1})$ is  the likelihood that random-angle samples have a weighted rms angle in excess of $w_{1}.$  

Note that, to a good approximation, the function $S(w)$ is a simple Gaussian, $S(w) \approx$ $e^{-w^2}.$ Nevertheless, the best fit formula (\ref{S}) is used in calculations. 

Following convention, the weighted rms angle for a sample is considered ``very significant'' when less than 1\% of random-angle samples have a larger value of $w.$ This occurs for $w \geq$ $w_{0}$ with $S(w_{0})$ = 0.01 or 
\begin{equation} \label{wVSig}
w \geq w_{0} = 2.252374 \quad \quad ({\mathrm{for}}\; \; S \leq  1\%)\, .
\end{equation}
With a way to judge the significance of the alignment of a sample, it is time to turn to evaluating the observed polarization data.

\section{Analysis of Data} \label{Results}

Of the 12746 radio sources in the catalog, 5442 sources have been selected  and bundled into 5350 samples for analysis, as discussed in Sec. \ref{data} above. The metric needed to judge the alignment of the sources in a sample is the weighted rms angle $w,$ which is calculated from the location of the sources in the sky and their observed polarization position angles via the process presented in Sec. \ref{GCT}. The significance of the alignment for a value of $w$ is given by the function $S(w),$ (\ref{S}). 

In this section, the great circle test for alignment in Sec. \ref{GCT} is applied to the 5350 samples consisting of the sources in $24^{\circ}$ radius circular regions centered on a $2^{\circ}$-spaced grid in the Northern Equatorial hemisphere. 

Calculations with observed quantities are complicated by uncertainties in the measured data. The 5442 sources  were chosen in part because their polarization angle uncertainties were less than $14^{\circ}.$ Uncertainties in location are considered small enough to ignore. To estimate the uncertainty in calculated results the calculations were run for 17 sets of data, one run with the best values of position angle $\theta_{p}$ and 16 runs with position angles $\theta_{p \, i}$ selected randomly in the indicated range of uncertainty $\theta_{p} - \sigma_{\theta} \leq$ $\theta_{p \, i} \leq$ $\theta_{p}+\sigma_{\theta},$ $i \in$ $\{1, ... ,16\}.$  The 16 runs produce 16 values $b_{i}$ for each calculated result, yielding 
\begin{equation} \label{uncertainR}
\bar{b} = \frac{1}{16}\sum_{i=1}^{16}{b_{i}} \quad ; \quad \sigma_{b} = \sqrt{\frac{1}{16}\left( \bar{b}-b_{i}  \right)^{2}} \, ,
\end{equation}  
for the mean $\bar{b}$ and uncertainty $\sigma_{b}.$

It can happen that the mean $\bar{b}$ and the best value $b$ are not close in value. In that case the best value $b$ is presented  in the text and some discussion is included if the quantity has some importance. When there is little difference between the mean $\bar{b}$ and the best value $b$ compared to $\sigma,$ the value presented in the text is $b \pm \sigma_{b},$ so we default to the best value of $b$ obtained with the values of $\theta_{p}$ presented in the catalog.

The quadrupole coefficients $a_{m}$ for each source are found by (\ref{am}) and then averaged over the sample to get the sample averages $\bar{a}_{m}.$ Next the coefficients $\bar{a}_{m}$ produce the symmetric matrix $T$ by (\ref{Tensor}). The eigensystem is solved and since the max-min eigenvalues are the max-min quadrupole angles $\bar{\eta}_{q \,\mathrm{max}}$ and $\bar{\eta}_{q \,\mathrm{min}},$ we can calculate the root-mean-square of these extremes, the rms angle $\bar{\eta}_{q \,\mathrm{rms}},$ one for each of the 5350 samples. Almost done. 

To account for the natural tendency of smaller samples to have larger extreme rms angles, the rms angle of a sample with $N$ sources is weighted using (\ref{w}); $w$ is normalized so that $w$ =1 for random-angle samples with any number $N$ sources. The significance of the 5350 $w$s is determined by  the significance function $S(w),$ (\ref{S}).  A `very significant' weighted rms angle $w$ occurs for less than 1\% of random-angle samples and that happens for $S(w)$ = 0.01 at $w$ = 2.252374. Any sample with $w$ = 2.252374 or larger is deemed very significant.


Of the 5350 samples, there were 158 very significant samples with $w$s ranging from $w$ = $2.2524$ up to $w$ = $2.974 \pm 0.089$, ranging in significance from 1\% down to $0.023\% \pm 0.015\%$ which means that less than 1 of every $4300,$ where 4300 = 1/0.00023, random-angle samples would have $w$ as large as 2.974.   

\begin{figure}[ht]  
\centering
\vspace{0cm}
\hspace{0in}\includegraphics[0,0][360,360]{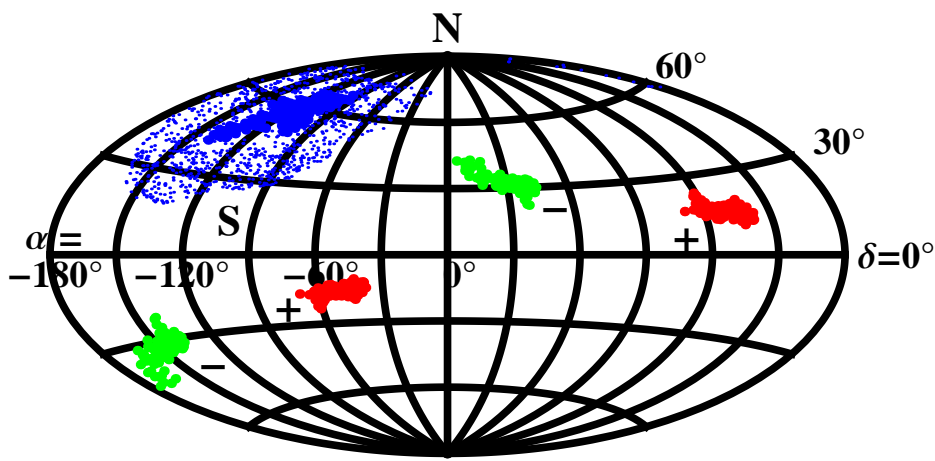}
\caption{(Color online) {\it{The cluster of 90 very significant samples and its $\bar{\eta}_{q \, {\mathrm{max}}}$ and $\bar{\eta}_{q\, {\mathrm{min}}}$ regions.}} The locations of the sources (S) are marked with small dots surrounding the clump of sample centers.  The polarization vectors of each sample favor the locations with minimum rms angle $\bar{\eta}_{q\, {\mathrm{min}}},$ labeled `$-$' and they tend to point away from the max rms angle regions, labeled `+'.  The three radial directions to a sample in the clump (S) and one of its maxima (+) and one of its minima ($-$) are nearly mutually orthogonal, forming a preferred rectangular coordinate system.   }
\label{90Cluster}
\end{figure}

\begin{figure}[ht]   
\centering
\vspace{0cm}
\hspace{0in}
\includegraphics[0,0][360,360]{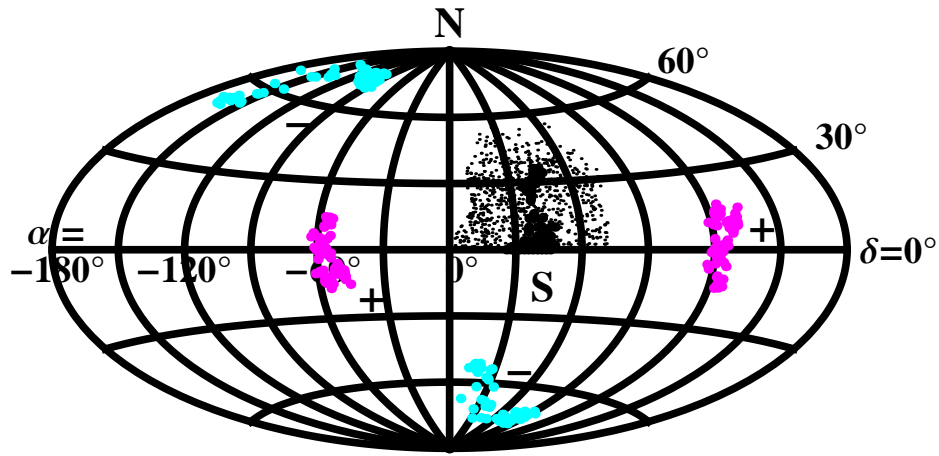}
\caption{(Color online) {\it{The cluster of 57 very significant samples and its $\bar{\eta}_{q \, {\mathrm{max}}}$ and $\bar{\eta}_{q\, {\mathrm{min}}}$ regions.}} As in Fig. \ref{90Cluster} five regions are identified. The locations of sources is labeled `S'. The two diametrically opposite clumps at the location of the minimum rms angle, labeled `$-$', indicate the directions of the best alignment of the samples.  Similarly a `+' is placed at the clumps of max rms angle locations, indicating the worst alignment of the samples' polarization vectors. As with the 90-sample cluster, the directions to one of the 57 sample centers at S and to either of its best alignment locations ($-$) and  to either of its worst alignment locations (+) are mutually perpendicular.  }
\label{57Cluster}
\end{figure}

\begin{figure}[ht]  
\centering
\vspace{0cm}
\hspace{0in}\includegraphics[0,0][360,360]{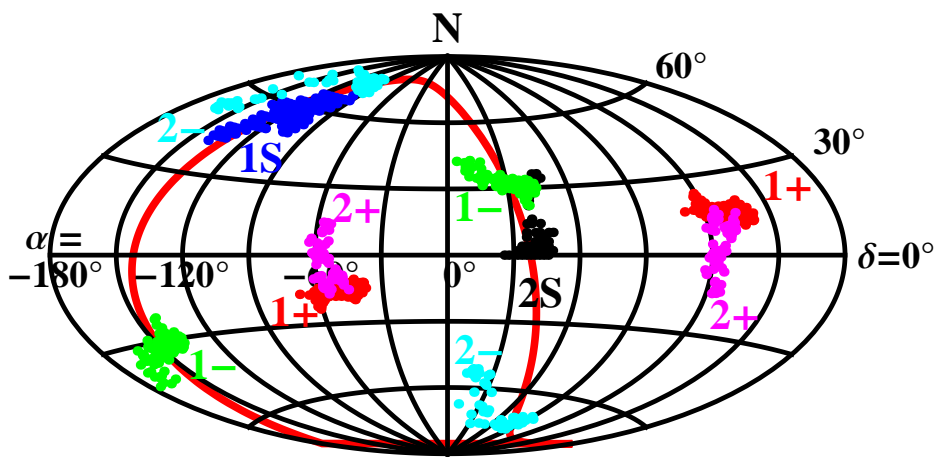}
\caption{(Color online) {\it{The best ($-$) and worst (+) alignment regions and the location (S) of the samples for both the 90- and the 57-sample clusters.}} This plot combines Figs. \ref{90Cluster} and \ref{57Cluster} except the clouds of sources are dropped and sample centers are kept. A `1' indicates the 90-sample cluster and `2' indicates the 57-sample cluster.  Many coincidences are apparent. The `1+' and `2+' regions of the 90- and 57- sample clusters are coincident with average locations separated by some $18^{\circ}$. The `$2-$' region of the 57-sample cluster is coincident with the 90-sample sources marked with `1S';  they are about $22^{\circ}$ apart. And the `$1-$' region of the 90-sample cluster is coincident with the 57-sample sources 2S, about $27^{\circ}$ apart. A preferred spherical coordinate system is inferred. The `1+` and `2+' regions inhabit the poles and the six remaining regions are laced along the equator of the preferred system. The equator for the inferred orientation is the solid meandering line (red online).      }
\label{TwoClusters}
\end{figure}

 The 158 very significant samples are located in four clusters on the sky: $90 \pm 16$ samples with centers near (RA,dec) $\approx$ $(-119^{\circ},57^{\circ}) ,$  $57 \pm 20$ samples near (RA,dec) $\approx$ $(40^{\circ},12^{\circ}),$ three samples at (RA,dec) $\approx$ $(-80^{\circ},3^{\circ}),$  and eight samples at (RA,dec) $\approx$ $(-40^{\circ},4^{\circ}),$ with RAs and decs to a couple of degrees or less.  The 3- and 8-sample clusters occur along the equator at the edge of the data and are ignored in what follows. We have two clusters to investigate, a 90- and a 57-sample cluster. See Figs. \ref{90Cluster}, \ref{57Cluster} and \ref{TwoClusters}.
 
The polarization vectors of a sample tend to point to the location in the sky where the rms angle is a minimum. The minimum rms angle is an eigenvalue of the matrix $T$ and the direction to the minimum is an eigenvector, as described in Sec. \ref{GCT}. Similarly for the maximum, the direction avoided by the sample's polarization vectors.  The directions to minima and maxima are plotted as well in Figs. \ref{90Cluster}, \ref{57Cluster} and \ref{TwoClusters}.  
 
For each sample, the $3\times3$ matrix $T$ has three distinct eigenvalue/eigenvector combinations. As noted in Sec. \ref{GCT}, the middle eigenvector, the eigenvector $\hat{r}_{\mathrm{mid}}$ for the mid-sized eigenvalue, ends up being close to the location of the sample. For the very significant samples, one finds that the average angle between the eigenvector $\hat{r}_{\mathrm{mid}}$ and the center of the sample region is $12.1^{\circ}\pm 1.3^{\circ}$ for the 90-sample cluster and $16.7^{\circ} \pm 2.2^{\circ} $ for the 57-sample cluster. The locations of $\hat{r}_{\mathrm{mid}}$ are not plotted to avoid clutter. 

The eigenvectors of distinct eigenvalues of $T$ are perpendicular.\cite{eigenvalues} This implies that the cluster of sample locations, the directions of preferred alignment, and the directions avoided by the polarization vectors are mutually perpendicular, as can be inferred from Figs. \ref{90Cluster} and~\ref{57Cluster}.

There are some coincidences evident in the figures. First note that in Fig. \ref{TwoClusters} the region of the sky avoided by the 90-sample cluster is coincident with the region avoided by the 57-sample cluster. The average position of the two clusters in the sky are separated by just $18.1^{\circ} \pm 2.8^{\circ},$ and the avoidance regions are spread out enough so that they overlap in parts. In Fig. \ref{TwoClusters} these are the two clumps of dots near RA = $-60^{\circ}$ on the Celestial equator and the diametrically opposite location near RA = $+120^{\circ}.$ 

As an immediate consequence of the eigenvectors being orthogonal, the coincidence of avoidance regions (`+' in the figures) forces the alignment regions (`$-$') and the sample centers (`S') into the same plane, along a single great circle in the sky. This great circle is in a plane in space, but on the Aitoff projection Fig. \ref{TwoClusters}, it is the solid line meandering about. Every clump of dots labeled `$-$' or `S' is connected by the great circle.  The `+' clumps are situated as `poles' with the solid line great circle as an `equator' in Figure \ref{TwoClusters}.
 
Another coincidence is that the sources (`S') and alignment regions (`$-$') are apparently not randomly dispersed along the solid line of the great circle in Fig. \ref{TwoClusters}. Instead, one of the alignment clump of the 90-samples is coincident with the location of the centers of the 57-samples and vica versa. The angle between the average location in the alignment clump of the 90-samples and the sample centers of the 57-sample cluster is $28.1^{\circ} \pm 1.6^{\circ}.$ And, vica versa, the angle between the location of the alignment clump of the 57-samples and the source region of the 90-samples is $21.9^{\circ} \pm 1.5^{\circ}.$ In addition, the clumps are spread out enough to overlap.

One must not assign a lot of importance to the coincidences found. It may be that they indicate some property of the medium traveled by the radio waves from source to detector. Or maybe not. It must be noted that such coincidences were not expected as part of the hypothesis looking for alignment that prompted the analysis. One cannot consider these as additional evidence in support of alignment. Nevertheless, the coincidences are there and worth noting, if only as something to look for in the analysis of new data.

The search for signs of alignment has turned up 158 very significant samples.  The most significant, the sample with $w$ = 2.9737, would be likely for less than 1 out of 4300 random-angled samples. If this sample was the only sample of data analyzed, then the result would be very significant indeed.

But there are 5350 samples on half the sky and one must ask for the likelihood that 5350 samples of random-angled sources would include 158 samples that are very significant. To answer this we can look at the 100 runs with random-angle data. Each of these runs analyzes the same 5350 samples but with a different set of random angles in place of the polarization angles of the 5442 sources. One finds that 7 of the 100 runs has more than 158 very significant samples in a batch of 5350 samples. 

The median number of very significant samples is 32, meaning that 50 runs had more very significant samples and 49 had fewer.  The observed data produces 158 samples, far more than the median of 32.

\begin{figure}[ht]  
\centering
\vspace{0cm}
\hspace{0in}\includegraphics[0,0][360,360]{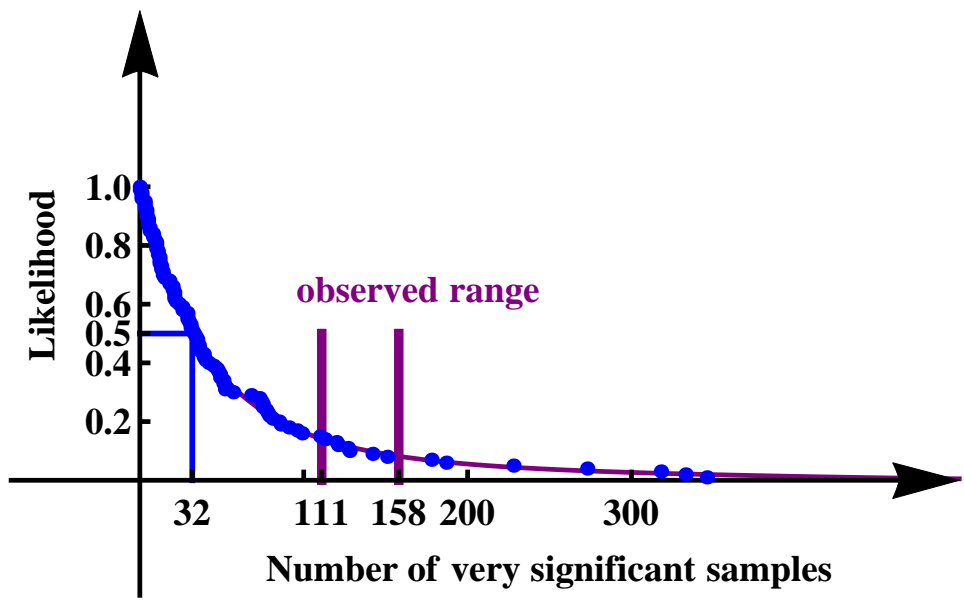}
\caption{(Color online) {\it{Distribution of the number of very significant samples in 100 random runs.}} The analysis applied once to the observed data was also applied a hundred more times to random data, the same sources but with random polarization  angles. The likelihood is the fraction of runs with more very significant samples.  Uncertainties in the data carry through to uncertainties in results so that a range of between 111 and 158 very significant samples were found. Compare that with the median run, the one with a 50\% likelihood, which had 32 very significant samples. Fourteen runs had more than 111 very significant samples and seven had more than 158, so the likelihood of a random run having 111 to 158 or more very significant samples is 7\% to 14\%. Thus, the likelihood that the alignment found for the observed data would be replicated by random data is about 7\% to 14\%.   }
\label{100RUNS}
\end{figure}

Complicating the situation are the uncertainties in the polarization vector position angles. The best position angles, the  catalog values $\theta_{p},$ yield the best value of 158 very significant samples. But, for the 16 runs with position angles $\theta_{p \, i} $ allowed by the uncertainty range, $ \theta_{p}- \sigma_{\theta} \leq \theta_{p \, i} \leq  \theta_{p}+ \sigma_{\theta}, $ the mean value is 136 very significant samples. And the rms difference with the mean is $\sigma$ = 25, so as few as $136 - 25$ = 111 very significant samples would be reasonably allowed by uncertainties in polarization angle. There were 14 of the 100 runs that had more than 111 very significant samples, so that 14\% is the fraction of random runs that would have at least 111 very significant samples. See Fig. \ref{100RUNS}.

One concludes that the level of alignment found by the great circle test would be found for about 7\% to 14\% of random-angle sources.  This suggests that these alignments may not be random and that there may be a physical process or processes responsible for the alignments of the from 111 to as many as 158 very significant samples. If the processes exist, then knowing the locations of the sources and the directions that their polarization vectors avoid and the directions they align with should assist in the investigation.

\section{Figures} \label{figs}

\end{document}